\documentclass{iaus}
\usepackage{graphicx}

\title[Bar/Bulge/Disk Decomposition]
{Multi-Band Bar/Bulge/Disk Image Decomposition of a Thousand Galaxies}

\author[Gadotti \& Kauffmann]
{Dimitri Gadotti \and Guinevere Kauffmann}

\affiliation{Max-Planck-Institut f\"ur Astrophysik \break
Karl-Schwarzschild-Str. 1, D-85741, Garching bei M\"unchen, Germany \break
e-mail: dimitri,gamk@mpa-garching.mpg.de}

\pubyear{2007}
\volume{241}
\jname{Stellar Populations as Building Blocks of Galaxies}
\editors{}
\begin{document}

\maketitle

\begin{abstract}
The structural parameters of bulges, disks and bars of a sample of nearly 1000 nearby galaxies are being determined
through sophisticated image decomposition in the $g$, $r$ and $i$ bands. The sample is carefully drawn from the Sloan Digital
Sky Survey Data Release 2 (SDSS DR2), contains 963 galaxies, of which 407 host AGN, and we show that it is
representative of the galaxy and AGN populations in the local universe. A large number of other physical
properties have also been determined for these galaxies within the SDSS collaboration. With these data, we reinforce
several recent studies and find a number of results leading to new insights into how the different galaxy components
relate to each other and the extent to which the hosts galaxies of AGN differ from quiescent galaxies.
\keywords{galaxies: active, galaxies: photometry, galaxies: structure}
\end{abstract}

\firstsection 

\section{Sample and Image Decomposition}

The galaxies in our sample are at $0.02\leq z\leq0.07$ and have $M_\star\geq10^{10}$ M$_{\odot}$.
In order to obtain most reliable results, our sample was selected to be suitable for image decomposition: galaxies
are close to face-on ($b/a\geq0.9$), are not morphologically disturbed, and have an apparent diameter larger than $\sim8''$.
The fits are done with the new version of {\sc budda} (\cite{des04}) and checked individually
to avoid wrong results that can arise from automated procedures. The models include up to four components: bulge,
disk, bar and central source. This allows fitting galaxies with bars without compromising bulge and disk parameters,
and to take into account the contribution from type 1 AGN. Figure 1 shows an illustrative example.

\begin{figure}
\includegraphics[width=6.75cm]{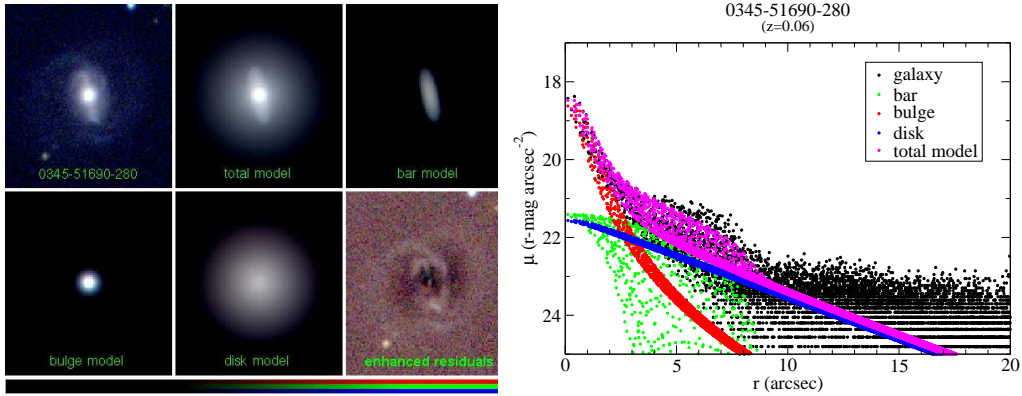}\includegraphics[width=6.75cm]{gadotti_fig1b.eps}\vskip 0.15cm
\caption{Example of the results obtained with {\sc budda} on one of the SDSS galaxies at $z=0.06$. At left we show $gri$
color composites of the original galaxy image, the total model, the models of each
separate component and a enhanced residual image. In the latter, one clearly sees the spiral arms and the inner
ring/lens fragments that surrounds the bar. The negative residuals also delineate the fainter inter-arm areas of
the disk. This is likewise true for the region between the bar and the inner ring. The highest residuals reach a
level of about 10\%. The radial profiles on the right also demonstrate the fitting ability of the code. Each
point in these profiles refer to a single pixel. Only a fraction of the pixels is shown.}
\end{figure}

\section{First Results}

Amongst our first main results, with $\sim1/3$ of the sample analysed, we find that:

\begin{itemize}

\item there is a correlation between the effective radius of the bulge and the 4000 \AA\ break index D$_{\rm n}$(4000), in the sense that
smaller bulges have younger stellar populations (see also \cite{tho06});

\item the $g-i$ colours of bulges and bars are strongly correlated for galaxies of all bulge-to-total
luminosity ratios (B/T). In fact, bulges
and bars have similar colours. We are currently investigating whether this is a result of dust attenuation or colour gradients.
We also find a correlation between the colours of disks and bulges that shows the same triangular pattern noted by \cite{kau06}
using ultraviolet colours.
Blue bulges are almost always surrounded by blue disks, but galaxies with red bulges can have either blue or red disks. Finally,
the correlation between bar and disk colours is weak, even for late-type galaxies, with low values of B/T. Bars are usually
redder than their disks;

\item there is a gradual increase in B/T and stellar mass from star forming galaxies (B/T $\sim0.14$) to composite systems
(B/T $\sim0.21$) and AGN hosts (B/T $\sim0.31$);

\item the median {\em bar}-to-total luminosity ratio is Bar/T $\sim0.1$ regardless of AGN or star formation activity and the total stellar
mass of the galaxy; Bar/T ranges from $\approx$ 0.01 to $\approx$ 0.3;

\item there is an anti-correlation between the effective radius of the bulge and the accretion rate onto the supermassive black hole
in AGN hosts, parameterized as $\log{\rm L[OIII]}/{\rm M}_\bullet$, in the sense that the hosts of the most powerful AGN have smaller bulges
(see also \cite{hec04}).
In addition, at a fixed bulge size, the hosts of the most powerful AGN have bluer disks, in consonance with the results in \cite{kau06};

\item the effective radius of the bulge is the {\em only} structural parameter that shows a clear correlation with the accretion
rate onto the black hole. The bulge S\'ersic index and the bulge effective surface brightness show only weak trends.
The accretion rate does not depend on any structural parameter of either the disk or the bar.

\end{itemize}

\begin{acknowledgments}
This work is supported by the Deutsche Forschungsgemeinschaft priority program 1177 (``Witnesses of Cosmic History: Formation and evolution,
of galaxies, black holes and their environment''), and the Max Planck Society.
\end{acknowledgments}

\end{document}